\documentclass[sigconf,screen,nonacm]{acmart}

\AtBeginDocument{%
  \providecommand\BibTeX{{%
    \normalfont B\kern-0.5em{\scshape i\kern-0.25em b}\kern-0.8em\TeX}}}

\setcopyright{none}
\copyrightyear{2022}
\acmYear{2022}
\acmDOI{XXXXXXX.XXXXXXX}

\usepackage{subfigure}
\usepackage{multirow}
\usepackage{xcolor}

\begin{document}

\title{ Realistic soft-body tearing under 10ms in VR }

\author{Manos Kamarianakis}
\email{kamarianakis@uoc.gr}
\orcid{0000-0001-6577-0354}
\affiliation{%
  \institution{FORTH - ICS, University of Crete, ORamaVR}
  \country{Greece}
}

\author{Antonis Protopsaltis}
\orcid{0000-0002-5670-1151}
\email{aprotopsaltis@uowm.gr}
\affiliation{%
  \institution{University of Western Macedonia, ORamaVR}
  \country{Greece}
}

\author{Michail Tamiolakis}
\orcid{0000-0002-9393-3138}
\email{michalis.tamiolakis@oramavr.com}
\affiliation{%
  \institution{University of Crete, ORamaVR}
  \country{Greece}
}
\author{George Papagiannakis}
\orcid{0000-0002-2977-9850}
\email{papagian@ics.forth.gr}
\affiliation{%
  \institution{FORTH - ICS, University of Crete, ORamaVR}
  \country{Greece}
}

\renewcommand{\shortauthors}{Kamarianakis, Protopsaltis,  Tamiolakis, et al.}


\begin{teaserfigure}
\centering
  \includegraphics[width=1\textwidth]{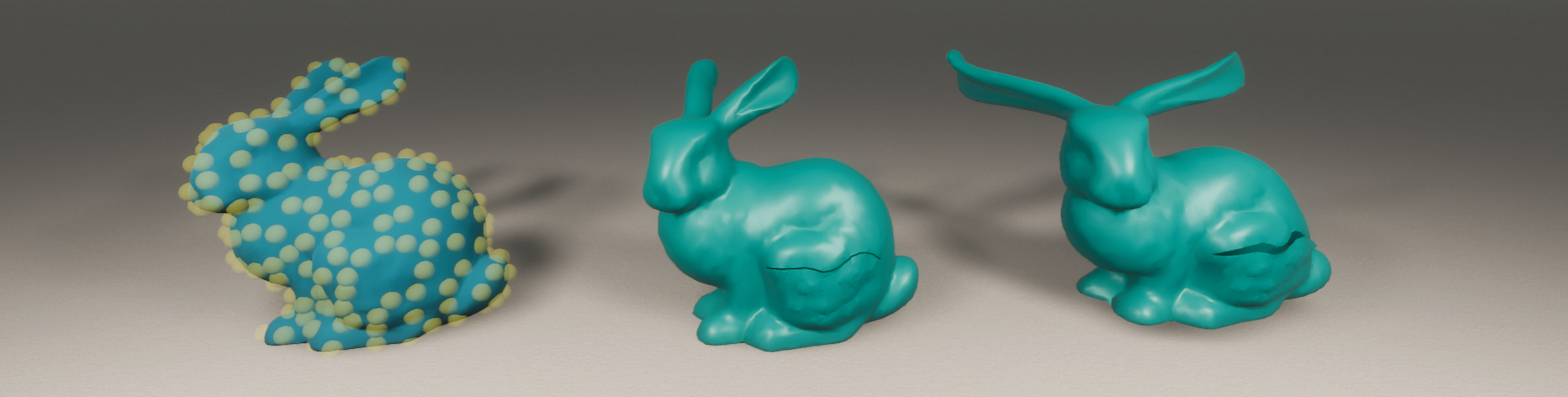}
  \caption{(Left) The vertices of a 3D  model are clustered in particles, to allow soft-body characteristics. (Middle) A continuous tear is 
  performed on the model. (Right) The particles of the torn model are updated, allowing further proper soft-body deformations.}
  \Description{A continuous tear is performed on 
  a bunny 3D model. The original and torn model 
  allow soft-body deformations, based on a clustering 
  of the vertices.}
  \label{fig:teaser}
\end{teaserfigure}

\maketitle

\section{Introduction}

Rigged animated models are one of the most researched areas in computer 
graphics and have been vastly adopted in  
Virtual Reality (VR) applications. VR experts experiment with various animation
and deformation techniques that can yield realistic real-time outputs. 
To cover the needs that arise from a variety of use cases, 
our research revolves around the ability to perform realistic tears, 
i.e., small cuts, on the surface of a model. 
Current bibliography \cite{Bruyns:2002jc, wu_efficient_2013} describes 
diverse ways on 
how to cut a 3D model, but most of these methods are not suitable for 
VR, since the specific calculations must be performed in a real-time 
manner within a few ms to preserve user immersion. Furthermore, to avoid 
the uncanny valley in VR, we emulate realistic effects while performing 
cuts on certain materials, such as a sponge or human tissues, that one 
would expect to occur in real life.

Latest developments \cite{Kamarianakis_Papagiannakis_2021} allow for 
basic operations, such as cutting, tearing or drilling on a rigged 
mesh model, to be run in near real-time. 
Furthermore, the replication of the physical behaviour of soft bodies, 
when applying external forces to them in VR, would greatly 
increase the overall realism of the simulation
\cite{macklin2014unified}.
The ongoing research for increased realism in virtual 
environments heavily impacts educational-oriented
applications, especially the ones regarding VR medical training
\cite{papagiannakisEditorialNewVirtual2022}.

\section{Our Approach}

\textbf{Overview.} Our approach is based on the techniques of
\cite{Kamarianakis_Papagiannakis_2021}, where the authors 
describe simple cut, tear and drill  operations on a 
3D mesh using basic geometric operations. Our optimized 
tearing module allows for continuous uninterrupted 
operation, i.e., the user can freely perform tears successively, 
similar to a surgeon's tearing gesture. 
Furthermore, we have developed a suitable particle decomposition on 
the model's vertices that can be used to emulate 
soft bodies, as in \cite{nealen2006physically}. 
Via suitable optimizations, we are able to perform real-time continuous 
tears on a soft-body model and update the underlying particle 
decomposition to obtain high-realistic results in VR. 
Our methods were designed with the lowest possible complexity in 
terms of needed calculations that yield real-time results even in 
untethered Head-Mounted Displays (HMDs),  with  limited 
GPU and CPU capabilities.
Lastly, proper handling and weight assignment 
\cite{Kamarianakis_Papagiannakis_2021} to the tear-generated vertices 
allow us to tear not only rigid but also skinned models, where in the 
latter case, further animation is still feasible.

\begin{figure}[htbp]
  \centering
  \includegraphics[width=0.45\textwidth]{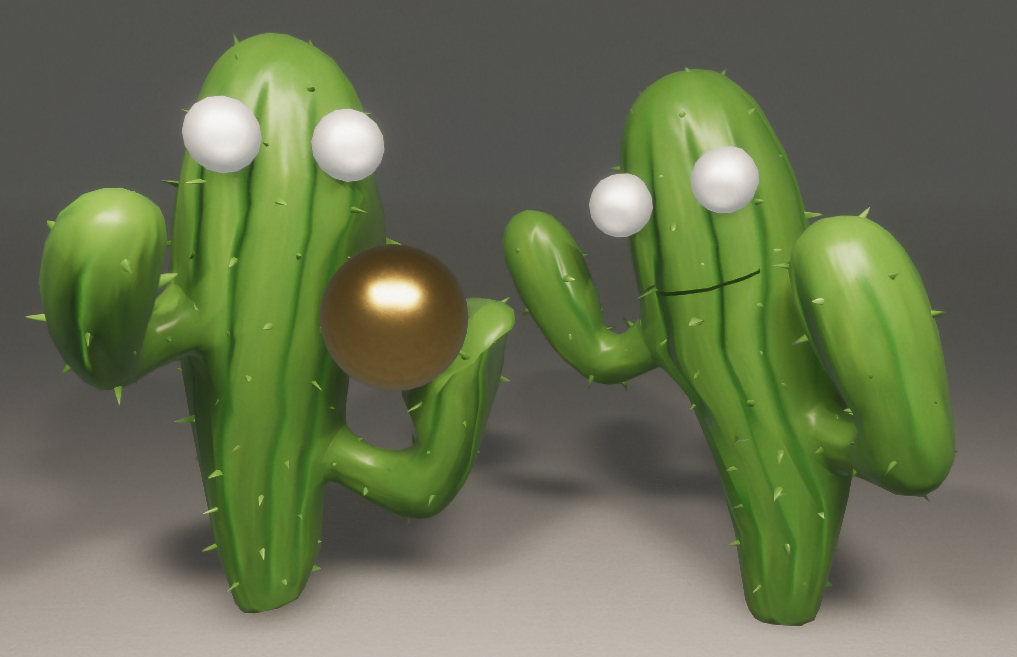}
  \Description{A cactus model with soft-bodies deformations enabled is torn. 
  The torn model can be further redeformed and still maintain soft-body
  characteristics.}
  \caption{Our methods applied on a cactus rigged model. Using particles we 
  can simulate soft-body simulations, as shown in the cactus holding the ball. 
  The skinned model can be torn and redeformed as shown on the right cactus. 
  The model's 2976 vertices were clustered into 110 particles, using $d=0.1$.}
  \label{fig:cactus}
\end{figure}

\textbf{Methodology for the Tear algorithm}
In order to achieve real-time tearing results, we have opted for basic geometric
primitives, e.g., face-plane intersections and face ray casting, as basic 
building blocks for our algorithms. 
This  approach allows for fast  identification of the faces affected 
by the tear. 

In our implementation, the tear width is user defined. In non-zero settings, 
a destructive tear takes place: triangles that fall in the tear-gap are 
completely or partially \emph{clipped}, i.e., removed from the model. 
Partially clipped faces are calculated by their intersections with the 
tear-gap surrounding box which initially is considered as a
finite part of a plane, defined and bounded by the intersections of the 
scalpel's initial and final positions with the model.

As the user moves the scalpel, freely tearing the model, we sample the 
scalpel's positions at specific time or distance intervals and perform multiple 
consecutive tears. 
To avoid jagged edges on the tearing path 
we make sure that consecutive bounding boxes do not overlap by utilizing  
non axis-aligned boxes instead. 

\balance
\textbf{Methodology for the Particle Decomposition}
To accomplish the so-called soft-body mesh deformation \cite{mages3}, 
the vertices of the mesh are
clustered into groups, called \emph{particles}. Each particle is 
positioned at the centroid of each group of vertices. A vertex may belong 
to multiple particles. In our method, every particle contains vertices within 
a  model dependent, euclidean range of $d$ units. 
A small range results in more particles, 
hence more accurate deformations, but yields worse running times.
Since a high range will have the opposite effects, a balanced range
should be identified per model (see 
Figure~\ref{fig:cactus}).

When a user applies a force to a particle, its position changes 
and this movement affects the position of all vertices 
of the particle. The particle's velocity is 
changed, proportionally to the displacement, always pointing 
to its initial position (simulating elasticity); in the case of 
skinned models, the current pose is  also considered.
Physics calculations are natively handled by the employed game engine. 

After a tearing operation, the clustering map is updated by adding or 
removing vertices to the involved particles. To allow 
fast updates and produce physically correct deformation results, simple 
directives were also introduced, e.g., vertices belonging to opposite 
sides of a tear, although close enough, cannot belong to the same particle. 

\textbf{Results}
Our optimized methods run on top of UNITY3D game engine and are 
incorporated within the MAGES SDK \cite{mages3}, 
developed by ORamaVR, publicly available for free.

Applying our methods on the Stanford bunny (3365 vertices), the 
tearing times were varying between 7-16ms depending on the  region torn. 
The average time for tears as the ones shown in Figure \ref{fig:teaser}  
was 10ms. The results were obtained without employing GPU or parallel 
processing, using an Intel core i7 7700HQ at 2.8GHZ with an Nvidia GTX 
1050ti m (8GB RAM) graphics card and a Desktop VR - HTC Vive. 
The  model's vertices were clustered into 78 particles, using range 
$d=0.1$, by an offline process that took 456ms.

\textbf{Comparison with other methods} 
Methods accomplishing similar results are implemented in the discontinued 
Nvidia Flex physics engine (\url{developer.nvidia.com/flex}) and the 
Flex-based Unity3D plugin uFlex that has not been updated in the 
past 6 years (still in beta). These methods do not support rigged 
models in their particle decomposition. They support cloth tearing but 
not surgical-like tearings as the ones described in our work. 
In conclusion, our method remains the only active, game-engine compatible, 
solution.

\section{Conclusions}\label{sec:conclusions_future_work}

We have presented a novel integration 
of a real-time continuous tearing algorithm for 3D
meshes in VR along with a suitable 
particle decomposition that allows
soft-body deformations on both the original 
and the torn model. Our methods are based 
on simple geometric primitives and 
therefore are suitable even for untethered 
HMDs of low specifications. 
The proposed techniques are already 
implemented in the MAGES SDK, running on Unity3D, publicly available for free.

\begin{acks}\label{sec:acks}
The project was partially funded by the European Union’s 
Horizon 2020 research and innovation programme under grant agreements 
No 871793 (ACCORDION) and No 101016509 (CHARITY).
\end{acks}

\bibliographystyle{ACM-Reference-Format}
\bibliography{references}

\end{document}